\documentclass[aps,amsfonts,pra,twocolumn,showpacs]{revtex4}
\usepackage{epsfig,amsmath,amssymb,bm,epsf,graphics}
\def\Real{{\rm Re\,}}
\def\bra#1{\langle#1\vert}
\def\ket#1{\vert#1\rangle}
\def\ketbra#1{\vert#1\rangle\langle#1\vert}
\def\ipr#1#2{\langle#1\vert#2\rangle}

\def\Longarrow{\protect\@lra}
\def\@lra{\relbar\joinrel\relbar\joinrel\relbar\joinrel%
          \relbar\joinrel\rightarrow}
\def\wstate{{\rm W}}
\def\wtilde{{\widetilde{\rm W}}}

\def\coe#1{{({\rm co}\,E_{#1})}}
\begin{document}
\title{Geometric measure of entanglement 
and applications \\
to bipartite and multipartite quantum states}

\author{Tzu-Chieh Wei and Paul M. Goldbart}
\affiliation{Department of Physics, 
University of Illinois at Urbana-Champaign, 
1110 West Green Street, Urbana, Illinois 61801-3080, U.S.A.}

\date{July 30, 2003}
% \date{\today}

\begin{abstract}
The degree to which a pure quantum state is entangled can be 
characterized by the distance or angle to the nearest unentangled 
state.  This geometric measure of entanglement, already present 
in a number of settings (see Shimony~\cite{Shimony95} and 
Barnum and Linden~\cite{BarnumLinden01}), is explored 
for bipartite and multipartite pure and mixed states.  The measure is 
determined analytically for 
arbitrary two-qubit mixed states and for generalized Werner and 
isotropic states, and is also applied to certain multipartite mixed 
states. In particular, a detailed analysis is given for arbitrary mixtures
of three-qubit GHZ, W and inverted-W states. Along the way,
we point out connections of the geometric measure
of entanglement with entanglement witnesses and with the Hartree approximation
method.
\end{abstract}
\maketitle

\section{Introduction}
Only recently, after more than half a century of existence, has the 
notion of entanglement become recognized as central to quantum 
information processing~\cite{NielsenChuang00}.  As a result, the 
task of characterizing and quantifying entanglement has emerged as one 
of the prominent themes of quantum information theory.  There have 
been many achievements in this direction, primarily in the setting of
{\it bipartite\/} systems~\cite{Horodecki01}.  Among these, one highlight 
is Wootters' formula~\cite{Wootters98} for the entanglement of formation 
for two-qubit mixed states; others include corresponding 
results for highly symmetrical states of higher-dimensional 
systems~\cite{VollbrechtWerner01,TerhalVollbrecht00}. 
The issue of entanglement for {\it multipartite\/} states 
poses an even greater challenge, and there have been 
correspondingly fewer achievements: notable examples  
include applications of the 
relative entropy~\cite{VedralPlenio98}, 
negativity~\cite{ZyczkowskiWerner}, and 
Schmidt measure~\cite{EisertBriegel01}.

In this Paper, we present an attempt to face this challenge by 
developing and investigating a certain geometric measure of entanglement (GME), 
first introduced by Shimony~\cite{Shimony95} in the setting of bipartite 
pure states and generalized to the multipartite setting (via projection 
operators of various ranks) by Barnum and Linden~\cite{BarnumLinden01}.  
We begin by examining this geometric measure in pure-state settings 
and establishing a connection with entanglement witnesses,
and then extend the measure to mixed states, showing that it satisfies 
certain criteria required of good entanglement measures. 
We demonstrate that this geometric measure is no harder to compute than the 
entanglement of formation $E_{\rm F}$, and exemplify this fact  
by giving formulas corresponding to $E_{\rm F}$ for 
(i)~arbitrary two-qubit mixed, 
(ii)~generalized Werner, and 
(iii)~isotropic states. 
We conclude by applying the geometric entanglement measure to certain 
families of multipartite mixed states, for which we provide a practical 
method for computing entanglement, and illustrate this method via several 
examples. In particular, a detailed analysis is given for arbitrary mixture
of three-qubit GHZ, W, and inverted-W states.

It is not our intention to cast aspersions on exisiting approaches to 
entanglement; rather we simply wish to add one further element to 
the discussion. Our discussion focuses on quantifying multipartite entanglement in terms of a single number rather than characterizing it. 

   The structure of the paper is as follows. In Sec.~\ref{sec:Pure} we describe the basic geometric ideas on quantifying entanglement geometrically in
the setting of pure quantum states, and establish a connection
with the Hartree approximation method and entanglement witnesses. In 
Sec.~\ref{sec:Mixed} we extend the definition of the GME to mixed states,
and show that it is an entanglement monotone. In Sec.~\ref{sec:Analytic}
we examine the GME for several families of mixed states of bipartite
systems:
(i)~arbitrary two-qubit mixed, 
(ii)~generalized Werner and
(iii)~isotropic states in bipartite systems, as well as (iv)~certain mixtures
of multipartite symmetric states. In Sec.~\ref{sec:GWW} we give a detailed application of
the GME to arbitrary mixtures of three-qubit GHZ, W and inverted-W states.
In Sec.~\ref{sec:Conclude} we discuss some open questions and further directions.  In App.~\ref{app:VW} we briefly review the Vollbrecht-Werner
technique used in the present Paper.

\section{Basic geometric ideas and application to pure states}
\label{sec:Pure}
We begin 
with an examination of entangled {\it pure\/} states, and of how one might 
quantify their entanglement by making use of simple ideas of Hilbert space 
geometry.  Let us start by developing a quite general formulation, appropriate 
for multipartite systems comprising $n$ parts, in which each part can have 
a distinct Hilbert space.  Consider a general $n$-partite pure state
\begin{equation}
|\psi\rangle=\sum_{p_1\cdots p_n}\chi_{p_1p_2\cdots p_n}
|e_{p_1}^{(1)}e_{p_2}^{(2)}\cdots e_{p_n}^{(n)}\rangle.
\end{equation}
One can envisage a geometric definition of its entanglement 
content via the distance 
\begin{equation}
d=\min_{|\phi\rangle}
\Vert\,|\psi\rangle-|\phi\rangle\Vert
\end{equation}
between $\ket{\psi}$ and the nearest separable state $\ket{\phi}$
(or equivalently the angle between them). Here, 
$|\phi\rangle\equiv\otimes_{i=1}^n|\phi^{(i)}\rangle$ 
is an arbitrary separable (i.e., Hartree) $n$-partite pure state, 
the index $i=1\ldots n$ labels the parts, and 
\begin{equation}
|\phi^{(i)}\rangle\equiv
\sum_{p_i}c_{p_i}^{(i)}\,|e_{p_i}^{(i)}\rangle.
\end{equation}
It seems natural to assert that the more entangled a state is, 
the further away it will be from its best unentangled approximant 
(and the wider will be the angle between them). 

To actually find the nearest separable state, it is convenient to 
minimize, instead of $d$, the quantity 
\begin{equation}\Vert|\psi\rangle-|\phi\rangle\Vert^2,
\end{equation} 
subject to the constraint 
$\langle\phi|\phi\rangle=1$. 
In fact, in solving the resulting stationarity condition one may 
restrict one's attention to the subset of solutions $\ket{\phi}$
that obey the further condition that for each factor 
$\ket{\phi^{(i)}}$ 
one has 
$\ipr{\phi^{(i)}}{\phi^{(i)}}=1$.  
Thus, one arrives at the {\it nonlinear eigenproblem\/} for the 
stationary $\ket{\phi}$:
\def\phast{\phantom{\ast}} 
\begin{subequations}
\label{eqn:Eigen}
\begin{eqnarray}
\!\!\!\!\!\!\!\sum_{p_1\cdots\widehat{p_i}\cdots p_n}
\chi_{p_1p_2\cdots p_n}^*{c_{p_1}^{(1)\phast}}\cdots\widehat{c_{p_i}^{(i)\phast}}\cdots c_{p_n}^{(n)\phast}&\!=\!&
\Lambda\,{c_{p_i}^{(i)*}}, \\ 
\!\!\!\!\!\!\!\sum_{p_1\cdots\widehat{p_i}\cdots p_n}\chi_{p_1p_2\cdots p_n} {c_{p_1}^{(1)*}}\cdots\widehat{{c_{p_i}^{(i)*}}}\cdots {c_{p_n}^{(n)*}}&\!=\!&
\Lambda\,c_{p_i}^{(i)\phast}\,,
\end{eqnarray}
\end{subequations} 
where the eigenvalue $\Lambda$ is associated with the Lagrange 
multiplier enforcing the constraint 
$\ipr{\phi}{\phi}\!=\!1$, 
and the symbol \,\,$\widehat{}$\,\, denotes exclusion.  
In basis-independent form, Eqs.~(\ref{eqn:Eigen}) read
\begin{subequations}
\label{eqn:EigenForm}
\begin{eqnarray}
\langle\psi|\Big(\mathop{\otimes}_{j(\ne i)}^n|\phi^{(j)}\rangle\Big)
&=&\Lambda\bra{\phi^{(i)}}, \\
\Big(\mathop{\otimes}_{j(\ne i)}^n\langle\phi^{(j)}|\Big)|\psi\rangle
&=&\Lambda\ket{\phi^{(i)}}.
\end{eqnarray}
\end{subequations}
From Eqs.~(\ref{eqn:Eigen}) or (\ref{eqn:EigenForm}) one readily sees 
that the eigenvalues $\Lambda$ are real, in $[-1,1]$, 
and independent of the choice of the local basis 
$\big\{|e_{p_i}^{(i)}\rangle\big\}$. 
Hence, the spectrum  $\Lambda$ is the cosine of the angle 
between $|\psi\rangle$ and $\ket{\phi}$; 
the largest, $\Lambda_{\max}$, which we call the 
{\it entanglement eigenvalue\/}, corresponds to the closest 
separable state and is equal to the maximal overlap
\begin{equation}\Lambda_{\max}=\max_{\phi} ||\ipr{\phi}{\psi}||,
\end{equation}
where $\ket{\phi}$
is an arbitrary separable pure state.   

Although, in determining the closest separable state, we have used the 
squared distance between the states, there are alternative 
(basis-independent) candidates for entanglement measures which are
related to it in an elementary way: 
the distance, 
the sine, 
or the sine squared of the angle $\theta$ between them 
(with $\cos\theta\equiv\Real\ipr{\psi}{\phi}$). 
We shall adopt $E_{\sin^2}\equiv 1-\Lambda_{\max}^2$ as our
entanglement measure because, as we shall see, when generalizing 
$E_{\sin^2}$ to mixed states we have been able to show that it 
satisfies a set of criteria demanded of entanglement measures.  
We remark that determining the entanglement of $|\psi\rangle$ is 
equivalent to finding the Hartree approximation to the ground-state 
of the auxiliary Hamiltonian
${\cal H}\equiv-|\psi\rangle\langle\psi|$~\cite{ref:Mohit}.

In bipartite applications, the eigenproblem~(\ref{eqn:Eigen}) 
is in fact {\it linear\/}, and solving it is actually equivalent to 
finding the Schmidt decomposition~\cite{Shimony95}.  
Moreover, the entanglement eigenvalue is equal to the maximal 
Schmidt coefficient.  By constrast, for the case of three or more parts, 
the eigenproblem is a {\it nonlinear\/} one. As such,   
one can in general only address it directly, i.e., by determining the eigenvalues and 
eigenvectors simultaneously and numerically.  Yet, as we shall 
illustrate shortly, there do exist certain families of pure states whose 
entanglement eigenvalues can be determined analytically. 

\subsection{Illustrative examples}
Suppose we are already in possession of the Schmidt 
decompostion of some two-qubit pure state:  
\begin{equation}
|\psi\rangle=
 \sqrt{p}\,|00\rangle
+\sqrt{1-p}\,|11\rangle.
\end{equation}  
Then we can read off the entanglement eigenvalue:  
\begin{equation}
\Lambda_{\max}=
\max\{\sqrt{p},\sqrt{1-p}\}.
\end{equation} 
Now, recall~\cite{Wootters98} that the concurrence $C$ for this state 
is $2\sqrt{p(1-p)}$.  Hence, one has
\begin{eqnarray}
\label{eqn:LamConc}
\Lambda_{\max}^{2}=\frac{1}{2}\left(1+\sqrt{1-C^{2}}\right),
\end{eqnarray}
which holds for arbitrary two-qubit pure states.

The possession of symmetry by a state can alleviate the difficulty 
associated with solving the nonlinear eigenproblem.  To see this, 
consider a state 
\begin{equation}
|\psi\rangle=
\sum_{p_1\cdots p_n}\chi_{p_1p_2\cdots p_n}
|e_{p_1}^{(1)}e_{p_2}^{(2)}\cdots e_{p_n}^{(n)}\rangle
\end{equation}
that obeys the symmetry that the nonzero 
amplitudes $\chi$ are invariant under permutations. 
What we mean by this is that, regardless of the dimensions of the 
factor Hilbert spaces, the amplitudes are only nonzero when the 
indices take on the first $\nu$ values (or can be arranged to do so 
by appropriate relabeling of the basis in each factor) and, moreover, 
that these amplitudes are invariant under permutations of the parties, 
i.e., 
$\chi_{\sigma_1\sigma_2\cdots\sigma_n}=
 \chi_{p_1p_2\cdots p_n}$, 
where the $\sigma$'s are any permutation of the $p$'s. (This symmetry 
may be obscured by arbitrary local unitary transformations.)  
For such states, it seems reasonable to anticipate that the closest 
Hartree approximant retains this permutation symmetry.  Assuming this 
to be the case---and numerical experiments of ours support this 
assumption---in the task of determining the entanglement eigenvalue 
one can start with the Ansatz that the closest 
separable state has the form 
\begin{equation}
|\phi\rangle\equiv\otimes_{i=1}^n\left(\sum_j c_j|e_j^{(i)}\rangle\right),
\end{equation}
i.e., is expressed in terms of copies of a single factor state, 
for which $c^{(i)}_j=c_j$.  To obtain 
the entanglement eigenvalue it is thus only necessary to maximize 
${\rm Re}\,\langle\phi|\psi\rangle$ 
with respect to $\{c_j\}_{j=1}^{\nu}$, a simpler task than maximization 
over the $\sum_{i=1}^{n}d_{i}$ amplitudes of a
generic product state.

To illustrate this symmetry-induced simplification, we consider several 
examples involving 
permutation-invariant states, first restricting our attention to 
the case $\nu=2$. 
The most natural realizations are $n$-qubit systems. 
One can label these symmetric states according to the
number of $0$'s, as follows~\cite{Stockton}:
\begin{equation}
|S(n,k)\rangle\equiv \sqrt{\frac{k!(n-k)!}{n!}}
\sum_{\rm{\scriptstyle permutations}}
|\underbrace{0\cdots0}_{k}\underbrace{1\cdots1}_{n-k}\rangle.
\end{equation}
As the amplitudes are all positive, one can assume that the closest 
Hartree state is of the form
\begin{equation}
|\phi\rangle=\big(\sqrt{p}\,|0\rangle+\sqrt{1-p}\,|1\rangle\big)^{\otimes n},
\end{equation} 
for which the maximal overlap (w.r.t.~$p$) gives the entanglement
eigenvalue for $|{\rm S}(n,k)\rangle$:
\begin{eqnarray}
\label{eqn:Lambda}
\Lambda_{\max}(n,k)=
\sqrt{\frac{n!}{k!(n\!-\!k)!}}
\left(\frac{k}{n}\right)^{\frac{k}{2}}
{\left(\frac{n-k}{n}\right)}^{\frac{n\!-\!k}{2}}.
\end{eqnarray}
For fixed $n$, the minimum $\Lambda_{\max}$ 
(and hence the maximum entanglement) 
among the $\ket{{\rm S}(n,k)}$'s 
occurs for 
$k=n/2$ (for $n$ even) and 
$k=(n\pm1)/2$ (for $n$ odd). 
In fact, for fixed $n$ the general permutation-invariant state can 
be expressed as 
$\sum_k \alpha_k\ket{{\rm S}(n,k)}$ with $\sum_k |\alpha_k|^2=1$.
The entanglement of such states can be addressed via the 
strategy that we have been discussing, i.e.,
via the maximization of a function of (at most) three real parameters.
The simplest example is provided by the $n$GHZ state: 
\begin{equation}
|n{\rm GHZ}\rangle\equiv
\big(\ket{{\rm S}(n,0)}+\ket{{\rm S}(n,n)}\big)/\sqrt{2}.
\end{equation}
It is easy to show that (for all $n$)
$\Lambda_{\max}(n{\rm GHZ})=1/\sqrt{2}$ 
and 
$E_{\sin^2}=1/2$.

We now focus our attention on three-qubit settings. Of these, 
the states $\ket{S(3,0)}=\ket{000}$ and $\ket{S(3,3)}=\ket{111}$ 
are not entangled and are, respectively, the components of the 
the 3-GHZ state:
$\ket{{\rm GHZ}}\equiv
\big(\ket{000}+\ket{111})/\sqrt{2}$.
The states 
$\ket{{\rm S}(3,2)}$ 
and 
$\ket{{\rm S}(3,1)}$, 
denoted 
\begin{subequations}
\begin{eqnarray} 
\!\!\!\!\!\!\!|{\rm W}\rangle&\equiv&\ket{{\rm S}(3,2)} =
\big(\ket{001}+\ket{010}+\ket{100}\big)/\sqrt{3},\\ 
\!\!\!\!\!\!\!\ket{\widetilde{\rm W}}&\equiv&\ket{{\rm S}(3,1)}=
\big(\ket{110}+\ket{101}+\ket{011}\big)/\sqrt{3},
\end{eqnarray}
\end{subequations}
are equally entangled, having 
$\Lambda_{\max}=2/3$ and $E_{\sin^2}=5/9$.  

\begin{figure}[t]
\centerline{\psfig{figure=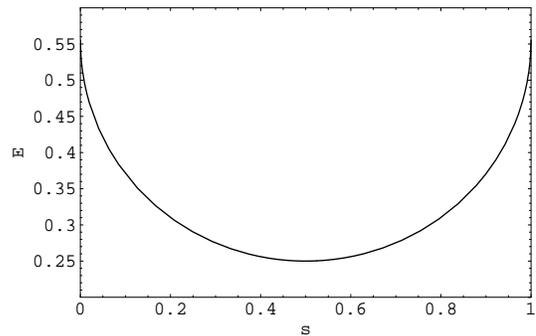,width=7cm,height=4.5cm,angle=0}}
%\vspace{- 0.4cm}
\caption{Entanglement of the pure state
$\sqrt{s}\,|{\rm W}\rangle+\sqrt{1-s}\,|\widetilde{\rm W}\rangle$ vs.~$s$. 
This also turns out to be the entanglement curve
for the mixed state $s\,\ketbra{\rm W}+(1-s)\ketbra{\widetilde{\rm W}}$.}
\label{fig:WW}
%\vspace{-0.5cm}
\end{figure}
Next, consider a superposition of the ${\rm W}$ and $\widetilde{\rm W}$ 
states: 
\begin{equation}
\ket{\wstate\wtilde(s,\phi)}
\equiv \sqrt{s}\,
\ket{\wstate}+\sqrt{1-s}\,{\rm e}^{i\phi}\ket{\wtilde}.
\end{equation}
It is easy to see that its entanglement is independent of
$\phi$: the transformation 
$\big\{\ket{0},\ket{1}\big\}\to
\big\{\ket{0},{\rm e}^{-i\phi}\ket{1}\big\}$
induces  
$\ket{\wstate\wtilde(s,\phi)}
\rightarrow 
{\rm e}^{-i\phi}\ket{\wstate\wtilde(s,0)}$.
To calculate $\Lambda_{\max}$, we assume that the 
separable state is 
$(\cos\theta\ket{0}+\sin\theta\ket{1})^{\otimes 3}$ 
and maximize its overlap with 
$\ket{\wstate\wtilde(s,0)}$.  
Thus we find that the tangent $t\equiv\tan\theta$ 
is the particular root of the polynomial equation
\begin{equation}
\sqrt{1-s}\,t^3+2\sqrt{s}\,t^2-2\sqrt{1-s}\,t-\sqrt{s}=0
\end{equation}
that lies in the range $t\in [\sqrt{{1}/{2}},\sqrt{2}]$.
Via $\theta(s)$, $\Lambda_{\max}$ 
(and thus $E_{\sin^2}=1\!-\!\Lambda_{\max}^2$) 
can be expressed as 
\begin{eqnarray}
\!\!\!\!\Lambda_{\max}(s)\!
=\!\frac{1}{2}\big[\sqrt{s}\cos\theta(s)\!
+\!\sqrt{1\!-\!s}\sin\theta(s)\big]
\sin2\theta(s).
\end{eqnarray}
In Fig.~\ref{fig:WW}, we show 
$E_{\sin^{2}}\big(\ket{\wstate\wtilde(s,\phi)}\big)$ vs.~$s$. 
In fact, $\Lambda_{\max}$ for the more general superposition   
\begin{equation}
\label{eqn:SSnk}
\ket{{\rm SS}_{n;k_1k_2}(r,\phi)}\equiv
\sqrt{r}\,                 \ket{{\rm S}(n,k_1)}+
\sqrt{1\!-\!r}\,{\rm e}^{i\phi}\,\ket{{\rm S}(n,k_2)}
\end{equation}
(with $k_1\ne k_2$) 
turns out to be independent of $\phi$, as in the case of 
$\ket{\wstate\wtilde(s,\phi)}$, and can be computed in the same way.
We note that although the curve in Fig.~\ref{fig:WW} is convex, 
convexity does not hold uniformly over $k_1$ and $k_2$. 

As our last pure-state example in the qubit setting, we consider 
superpositions of W and GHZ states: 
\begin{equation}
\ket{{\rm GW}(s,\phi)}
\equiv\sqrt{s}\,
\ket{{\rm GHZ}}+\sqrt{1-s}\,\,{\rm e}^{i\phi}\ket{{\rm W}}.
\end{equation}
For these, the phase $\phi$ cannot be ``gauged'' away and,
hence, $E_{\sin^2}$ depends on $\phi$. 
\begin{figure}[t]
\centerline{\psfig{figure=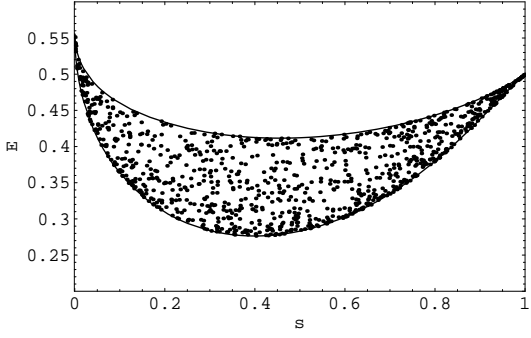,width=7cm,height=4.5cm,angle=0}}
%\vspace{- 0.4cm}
\caption{Entanglement of 
$\ket{{\rm GW}(s,\phi)}$ vs.~$s$.  
The upper curve is for $\phi=\pi$ whereas the lower one is for $\phi=0$. 
Dots represent states with randomly generated $s$ and $\phi$.}
\label{fig:WGHZ}
%\vspace{-0.5cm}
\end{figure}
In Fig.~\ref{fig:WGHZ} we show $E_{\sin^2}$ vs.~$s$ at $\phi=0$ and 
$\phi=\pi$ (i.e., the bounding curves), as well as $E_{\sin^2}$ for randomly generated 
values of $s\in[0,1]$ and $\phi\in[0,2\pi]$ (dots).  It is interesting 
to observe that the \lq$\pi$\rq\ state has higher entanglement than 
the \lq$0$\rq\ does.  As the numerical results suggest, the 
($\phi$-parametrized) $E_{\sin^2}$ vs.~$s$ curves of the states 
$\ket{{\rm GW}(s,\phi)}$ lie between the \lq$\pi$\rq\ and \lq$0$\rq\ curves.

We remark that, more generally, systems comprising $n$ parts, each a $d$-level
system, the symmetric state
\begin{equation}
\!\!\!\ket{S(n;\{k\})}\equiv \sqrt{\frac{\prod_i k_i!}{n!}}
\sum_{\rm{\scriptstyle P}}
|\underbrace{0..0}_{k_0}\underbrace{1..1}_{k_1}\dots\underbrace{(d\!-\!1)
.. (d\!-\!1)\,}_{k_{d\!-\!1}}\rangle,
\end{equation}
with $\sum_i k_i=n$, has entanglement eigenvalue 
\begin{equation}
\Lambda_{\max}(n;\{k\})=\sqrt{\frac{n!}{\prod_i (k_i!)}}
\,\prod_{i=0}^{d-1}\left(\frac{k_i}{n}\right)^{\frac{k_i}{2}}.
\end{equation}

One can also consider other symmetries. For example, for the 
totally antisymmetric (viz.~determinant) state of $n$ parts, each with $n$
levels,
\begin{equation}
\ket{{\rm Det}_n}\equiv \frac{1}{\sqrt{n!}}\sum_{i_1,\dots,i_n=1}^{n}
\epsilon_{i_1,\dots,i_n}\ket{i_1,\dots,i_n},
\end{equation}
it has been shown by Bravyi~\cite{Bravyi02} that the maximal squared overlap is $\Lambda^2_{\max}=1/n!$.
Bravyi also generalized the anti-symmetric state
to the $n=p\,d^p$-partite determinant state via
\begin{eqnarray}
\phi(1)&=&(0,0,\dots,0,0), \nonumber \\
\phi(2)&=&(0,0,\dots,0,1), \nonumber \\
&\vdots& \nonumber \\
\phi(d^p\!-\!1)&=&(d\!-\!1,d\!-\!1,\dots,d\!-\!1,d\!-\!2), \nonumber \\
\phi(d^p)&=&(d\!-\!1,d\!-\!1,\dots,d\!-\!1,d\!-\!1),\nonumber
\end{eqnarray}
and 
\begin{equation}
\ket{{\rm Det}_{n,d}}\equiv\frac{1}{\sqrt{(d^p!)}}\sum_{i_1,\dots,i_{d^p}}
\epsilon_{i_1,\dots,i_{d^p}}\ket{\phi(i_1),\dots,\phi(i_{d^p})}.
\end{equation}
In this case, one can show that $\Lambda^2_{\max}=[(d^p)!]^{-1}$.

\subsection{Connection with entanglement witnesses}
We now digress to discuss the relationship between the geometric measure
of entanglement and another entanglement property---entanglement
witnesses.
The entanglement witness ${\cal W}$ for an entangled state $\rho$ is
defined to be an operator that is (a)~Hermitian and (b)~obeys the following 
conditions~\cite{Horodecki396}:\\
 (i)~${\rm Tr}({\cal W}\sigma)\ge 0$ for all separable states $\sigma$, and\\
(ii)~${\rm Tr}({\cal W}\rho)< 0$.\\
Here, we wish to establish a correspondence between $\Lambda_{\max}$ for 
the entangled pure state $\ket{\psi}$ and the optimal element of the set 
of entanglement witnesses ${\cal W}$ for $\ket{\psi}$ that have the specific 
form 
\begin{equation}
\label{eqn:classW}
{\cal W}=\lambda^{2}\openone-\ketbra{\psi}, 
\end{equation} 
this set being parametrized by the real, non-negative number $\lambda^{2}$. By 
{\it optimal\/} we mean that, for this specific form of witnesses, 
the value of the \lq\lq detector\rq\rq\ 
${\rm Tr}\big({\cal W}\ketbra{\psi}\big)$ is as negative as can be. 

In order to satisfy condition~(i) we must ensure that, for any 
{\it separable\/} state $\sigma$, we have 
${\rm Tr}\big({\cal W}\sigma\big)\ge 0$.  
As the density matrix for any separable state can be decomposed 
into a mixture of {\it separable pure\/} states 
[i.e., $\sigma=\sum_i\ketbra{\phi_i}$ 
where $\{\ket{\phi_i}\}$ are separable pure states], 
condition~(i) will be satisfied as long as 
${\rm Tr}\big({\cal W}\ketbra{\phi}\big)\ge 0$ 
for all separable {\it pure} states $\ket{\phi}$.
This condition is equivalent to 
\begin{equation}
\lambda^{2} -||\ipr{\psi}{\phi}||^2\ge 0\ 
(\mbox{for all separable}\ \ket{\phi}),
\end{equation}
which leads to 
\begin{equation}
\lambda^{2} \ge
\max_{\ket{\phi}}||\ipr{\psi}{\phi}||^2=
\Lambda^2_{\max}(\ket{\psi}).
\end{equation}

Condition~(ii) requires that 
${\rm Tr}\big({\cal W}\ketbra{\psi}\big)<0$, in order for ${\cal W}$ 
to be a valid entanglement witness for $\ket{\psi}$; 
this gives $\lambda^{2} - 1<0$.  
Thus, we have established the range of $\lambda$ for which 
$\lambda^{2}\openone-\ketbra{\psi}$ 
is a valid entanglement witness for $\ket{\psi}$: 
\begin{equation}
\Lambda^2_{\max}(\ket{\psi}) \le \lambda^{2} < 1.
\end{equation}

With these preliminaries in place, we can now establish the connection 
we are seeking.  Of the specific family~(\ref{eqn:classW}) of 
entanglement witnesses for $\ket{\psi}$, 
the one of the form 
${\cal W}_{\rm opt}=\Lambda^2_{\max}(\ket{\psi})\openone -\ketbra{\psi}$ 
is optimal, in the sense that it achieves the most negative value for 
the detector 
${\rm Tr}\big({\cal W}_{\rm opt}\ketbra{\psi}\big)$: 
\begin{equation}
\min_{{\cal W}}{\rm Tr}\big({\cal W}\ketbra{\psi}\big)=
{\rm Tr}\big({\cal W}_{\rm opt}\ketbra{\psi}\big)= 
-E_{\sin^2}(\ket{\psi}),
\end{equation}
where ${\cal W}$ runs over the class~(\ref{eqn:classW}) 
of witnesses. 

We now look at some examples. For the GHZ state 
the optimal witness is 
\begin{equation}
{\cal W}_{\rm GHZ}=\frac{1}{2}\openone-\ketbra{\rm GHZ}
\end{equation}
and
${\rm Tr}\big({\cal W}_{\rm GHZ}\ketbra{{\rm GHZ}}\big)=
-E_{\sin^2}(\ket{{\rm GHZ}})=-1/2$. 
Similarly, for the W and inverted-W states 
we have 
\begin{equation}
{\cal W}_{\rm W}=\frac{4}{9}\openone-\ketbra{\rm W} \ \ \mbox{and} \ \
{\cal W}_{\rm \widetilde{\rm W}}=\frac{4}{9}\openone-\ketbra{\widetilde{\rm W}}
\end{equation}
and 
${\rm Tr}\big({\cal W}_{\rm W}\ketbra{{\rm W}}\big)=
-E_{\sin^2}(\ket{{\rm W}})=-5/9$, and similarly for $\ket{\widetilde{\rm W}}$. 
For the four-qubit state 
\begin{equation}
\ket{\Psi}\equiv
(\ket{0011}+\ket{0101}+\ket{0110}+
 \ket{1001}+\ket{1010}+\ket{1100})/\sqrt{6}
\end{equation}
the optimal witness is
\begin{equation}
{\cal W}_{\Psi}=\frac{3}{8}\openone-\ketbra{\Psi}
\end{equation}
and 
${\rm Tr}\big({\cal W}_{\Psi}\ketbra{\Psi}\big)=
-E_{\sin^2}(\ket{\Psi})=-5/8$.

Although the observations we have made in this section are, from a technical 
standpoint, elementary, we nevertheless find it intriguing that two distinct
aspects of entanglement---the geometric measure of entanglement and
entanglement witnesses---are so closely related.  Furthermore, this connection sheds 
new light on the content of the geometric measure of entanglement.  In 
particular, as entanglement witnesses are Hermitian operators, they can, at
least in principle, be realized and measured locally~\cite{Guhne}. Their connection with the
geometric measure of entanglement ensures that the geometric measure of
entanglement can, at least in principle, be verified experimentally.

\section{Extension to mixed states}
\label{sec:Mixed}
The extension of the GME to mixed states $\rho$ can be made via the use of the 
{\it convex roof\/} (or {\it hull\/}) construction [indicated by ``co''], 
as was done for the entanglement of formation 
(see, e.g., Ref.~\cite{Wootters98}).  The essence is a minimization 
over all decompositions $\rho=\sum_i p_i\,|\psi_i\rangle\langle\psi_i|$ 
into pure states, i.e., 
\begin{eqnarray}
\label{eqn:Emixed}
E(\rho)
\equiv
\coe{\rm pure}(\rho)
\equiv
{\min_{\{p_i,\psi_i\}}}
\sum\nolimits_i p_i \, 
E_{\rm pure}(|\psi_i\rangle).
\end{eqnarray}
Now, any good entanglement measure $E$ should, at least, 
satisfy the following criteria 
(c.f.~Refs.~\cite{VedralPlenio98,Horodecki300,Vidal00}):
\begin{itemize}
\item[C1.](a)~$E(\rho)\!\ge\! 0$; 
(b)~$E(\rho)\!=\!0$ if $\rho$ is not entangled.
\item[C2.]Local unitary transformations do not change $E$.
\item[C3.]Local operations and classical communication (LOCC)
(as well as post-selection) do not increase the 
expectation value of $E$.
\item[C4.]Entanglement is convex under the discarding of 
information, i.e., $\sum_i p_i\,E(\rho_i)\ge E(\sum_i p_i\,\rho_i)$. 
\end{itemize}
The issue of the desirability of additional features, such as 
continuity and additivity, requires further investigation, but 
C1-C4 are regarded as the minimal set, if one is to guarantee 
that one has an {\it entanglement monotone\/}~\cite{Vidal00}.  

Does the geometric measure of entanglement obey C1-4? 
From the definition~(\ref{eqn:Emixed}) it is evident that 
C1 and C2 are satisfied, provided that $E_{\rm pure}$ satisfies them, 
as it does for $E_{\rm pure}$ being any function of $\Lambda_{\max}$
consistent with C1.  It is straightforward to check that C4 holds,  
by the convex hull construction.  
The consideration of C3 seems to be more delicate. 
The reason is that our analysis of whether or not it holds 
depends on the explicit form of $E_{\rm pure}$. 
For C3 to hold, it is sufficient to show that the average entanglement is
non-increasing under any trace-preserving, unilocal operation: 
$\rho\rightarrow \sum_k V_k\rho V_k^\dagger$, where the Kraus
operator has the form  
$V_k=\openone\otimes\cdots \openone\otimes V_k^{(i)}\otimes\openone\cdots\otimes\openone$
and $\sum_k V_k^{\dagger} V_k=\openone$. Furthermore, it suffices to
show that C3 holds for the case of a pure  initial state, i.e., $\rho=\ketbra{\psi}$.  

We now prove that for the particular (and by no means un-natural) 
choice $E_{\rm pure}=E_{\sin^2}$, C3 holds. To be precise,
for any quantum operation 
on a pure initial state, i.e.,
\begin{equation}
|\psi\rangle\langle\psi|
\rightarrow\sum\nolimits_k V_k|\psi\rangle\langle\psi|V_k^\dagger,
\end{equation}
we aim to show that
\begin{equation}
\sum_k p_k \,E_{\sin^{2}}
\left({V_k|\psi\rangle}/\!{\sqrt{p_k}}\right)
\!\le\!
E_{\sin^{2}}(|\psi\rangle),
\end{equation}
where 
$p_k\!\equiv\!
{\rm Tr}\,V_k|\psi\rangle\langle\psi|V_k^\dagger
=\langle\psi|V_k^\dagger V_k|\psi\rangle$,
regardless of whether the operation $\{V_k\}$ 
is state-to-state or state-to-ensemble.
Let us respectively denote by $\Lambda$ and $\Lambda_k$ 
the entanglement eigenvalues corresponding to $\ket{\psi}$ 
and the (normalized) pure state ${V_k|\psi\rangle}/{\sqrt{p_k}}\,$. 
Then our task is to show that 
$\sum_k p_k\,\Lambda_k^2 \ge \Lambda^2$,
of which the left hand side is, by the definition of $\Lambda_k$, 
equivalent to 
\begin{equation}
\sum_k p_k 
\max_{\scriptscriptstyle\xi_k\in D_s}
\Vert
{\langle\xi_k| V_k|\psi\rangle}/\!{\sqrt{p_k}}
\Vert^2=\sum_k 
\max_{\scriptscriptstyle\xi_k\in D_s}\Vert
\langle\xi_k| V_k|
\psi\rangle\Vert^2.
\end{equation} 
Without loss of generality, we may assume that it is the first party
who performs the operation. Recall that the  
condition~(\ref{eqn:EigenForm}) for the closest separable 
state 
\begin{equation}
\ket{\phi}\equiv|\tilde{\alpha}\rangle_1\otimes|\tilde{\gamma}
\rangle_{2\cdots n}
\end{equation}
can be recast as
\begin{equation}
{}_{2\cdots n}\langle\tilde{\gamma}|\psi\rangle_{1\cdots n}=
\Lambda|\tilde{\alpha}\rangle_1.
\end{equation}
Then, by making the specific choice 
\begin{equation}
\langle\xi_k|=
({\langle\tilde{\alpha}|V_k^{(1)\dagger}/\!{\sqrt{q_k}}})
\otimes\langle\tilde{\gamma}|,
\end{equation}
where
$q_k
\equiv
\langle\tilde{\alpha}|
V_k^{(1)\dagger}V_k^{(1)}
|\tilde{\alpha}\rangle$,
we have the sought result
\begin{eqnarray}
&&\sum_k p_k\Lambda_k^2=
\sum_k \max_{\xi_k\in D_s}
\Vert\langle\xi_k| V_k|\psi\rangle\Vert^2\nonumber \\
&&\qquad\ge\Lambda^2\sum_k 
({\langle\tilde{\alpha}|
V_k^{(1)\dagger}V_k^{(1)}|
\tilde{\alpha}\rangle/\!\sqrt{q_k}}
)^2=\Lambda^2.
\end{eqnarray}
  Hence, the form $1-\Lambda^2$, when generalized to
mixed states, is an entanglement monotone. We note that a different approach to establishing this result has been used by 
Barnum and Linden~\cite{BarnumLinden01}.
Moreover, using the result that $\sum_k p_k\Lambda_k^2\ge \Lambda^2$,
one can further show that for any convex increasing function $f_c(x)$ with $x\in[0,1]$,
\begin{equation}
\sum_k p_k\,f_c(\Lambda_k^{2})\ge f_c(\Lambda^{2}).
\end{equation}
Therefore, the quantity
${const.}-f_c(\Lambda^{2})$ (where the ${const.}$ is to ensure the whole expression
is non-negative), when extended to mixed states, is
also an entanglement monotone, hence a good entanglement measure.
For the following discussion we simply take $E=1-\Lambda^2$.

\section{Analytic results for mixed states}
\label{sec:Analytic}
Before moving on to the {\it terra incognita\/} of mixed {\it multipartite\/}
entanglement, we test the geometric approach 
in the setting of mixed {\it bipartite\/} states, 
by computing $E_{\sin^2}$ for three classes of states for 
which $E_{\rm F}$ is known. 
\subsection{Arbitrary two-qubit mixed states}
For these we show that 
\begin{eqnarray}
\label{eqn:EC}
E_{\sin^2}(\rho)=
\frac{1}{2}\big(1-\sqrt{1-C(\rho)^2}\,\big),
\end{eqnarray}
where $C(\rho)$ is the Wootters concurrence of the state $\rho$. 
Recall that in his derivation of the formula for 
$E_{\rm F}$, Wootters showed that there exists 
an optimal decomposition $\rho=\sum_{i}p_i\,|\psi_i\rangle\langle\psi_i|$  
in which every $|\psi_i\rangle$ has the concurrence of $\rho$ itself.  
(More explicitly, every $|\psi_i\rangle$ has the identical  
concurrence, that concurrence being the infimum over all  
decompositions.)
By using Eq.~(\ref{eqn:LamConc}) one can, via Eq.~({\ref{eqn:EC}),
relate $E_{\sin^2}$ to $C$ for any two-qubit {\it pure\/} state. 
As $E_{\sin^2}$ is a monotonically increasing convex function of $C\in[0,1]$, 
the optimal decomposition for $E_{\sin^2}$ is identical to that for 
the entanglement of formation $E_{\rm F}$.  Thus, we see that 
Eq.~(\ref{eqn:EC}) holds for {\it any two-qubit mixed state}.

The fact that $E_{\sin^2}$ is  related to $E_{\rm F}$ via the concurrence $C$ is
inevitable for two-qubit systems, as both are fully determined by the one 
independent Schmidt coefficient.  
We note that Vidal~\cite{Vidalb00} has derived this expression when
he considered the probability of success for converting a single copy of
some pure state into the desired mixed state, which gives a physical
interpretation of the geometric measure of entanglement. Unfortunately, this
connection only holds for two-qubit states. 

\subsection{Generalized Werner states}
Any state $\rho_{\rm W}$ of a $C^d\otimes C^d$ system is called a generalized 
Werner state if it is invariant under
\begin{equation}
{\rm\bf P}_{1}:\rho\rightarrow
\int dU(U\otimes U)\rho\,
(U^\dagger\otimes U^\dagger),
\end{equation}
where $U$ is any element of the unitary group ${\cal U}(d)$ and $dU$ is 
the corresponding normalized Haar measure.  Such states 
can be expressed as a linear combination of two operators: 
the {\it identity\/} $\hat{\openone}$, and 
the {\it swap\/} $\hat{\rm F}\equiv\sum_{ij}|ij\rangle\langle ji|$, i.e., 
$\rho_{\rm W}\equiv a\hat{\openone}+b\hat{\rm F}$, 
where $a$ and $b$ are real parameters related via the 
constraint 
${\rm Tr}\rho_{\rm W}=1$. 
This one-parameter family of states can be neatly expressed in 
terms of the single parameter 
$f\equiv{\rm Tr}(\rho_{\rm W}\hat{\rm F})$:
\begin{equation}
\label{eqn:Wernerf}
\rho_{\rm W}(f)=\frac{d^2-fd}{d^4-d^2}\openone\otimes\openone+
\frac{fd^2-d}{d^4-d^2}\hat{\rm F}.
\end{equation}
By applying to $E_{\sin^2}$ the technique by developed by Vollbrecht 
and Werner 
for $E_{\rm F}(\rho_{\rm W})$ [see Ref.~\cite{VollbrechtWerner01} or App.~\ref{app:VW}], 
one arrives at
the geometric entanglement function for Werner states: 
\begin{eqnarray}
 \label{eqn:Werner}
 E_{\sin^2}\big(\rho_{\rm W}(f)\big)=
 \frac{1}{2}\big({1-\sqrt{1-f^2}}\,\big)\quad {\rm for} \ f\le 0,
 \end{eqnarray}
 and zero otherwise.

The essential points of the derivation are as follows:\\
(i)~In order to find the set $M_{\rho_{\rm W}}$ (see 
App.~\ref{app:VW}) it is sufficient, 
due to the invariance of $\rho_{\rm W}$ under ${\rm\bf P_1}$, to consider any pure state 
$|\Phi\rangle=\sum_{jk}\Phi_{jk}
|e^{(1)}_j\rangle\otimes|e^{(2)}_k\rangle$ that has 
a diagonal reduced density matrix 
${\rm Tr}_2\ketbra{\Phi}$ 
and the value ${\rm Tr}(|\Phi\rangle\langle\Phi|\hat{F})$ 
equal to the parameter $f$ associated with the Werner state 
$\rho_{\rm W}(f)$.  It can be shown that
\begin{equation}
E_{\sin^2}(|\Phi\rangle\langle\Phi|)
\ge\frac{1}{2}
\left({1-\sqrt{1-(f-\sum\nolimits_i\lambda_{ii})^2}}\,\right),
\end{equation}
where $\lambda_{ii}\equiv|\Phi_{ii}|^2$.
\\
(ii)~If $f>0$, we can set the only nonzero elements of $|\Phi\rangle$ 
to be 
$\Phi_{i1}$, 
$\Phi_{i2},\ldots$, 
$\Phi_{ii},\ldots$,
$\Phi_{id}$ such that
$|\Phi_{ii}|^2=f$, 
this state obviously being separable.  
Hence, for $f>0$ we have 
$E_{\sin^2}(\rho_{\rm W}(f))=0$.  
On the other hand, if $f<0$ then any nonzero $\lambda_{ii}$ 
would increase $(f-\sum_i\lambda_{ii})^2$ and, hence, 
increase the value of $E(|\Phi\rangle\langle\Phi|)$, not
conforming with the convex hull. 
Thus, for a fixed value of $f$, the lowest possible value of the 
entanglement $E(|\Phi\rangle\langle\Phi|)$ that can be achieved 
occurs when 
$\lambda_{ii}=0$ 
and there are only two nonzero elements 
$\Phi_{ij}$ and $\Phi_{ji}$ ($i\ne j$). 
This leads to 
\begin{equation}
\min_{\ket{\Phi} {\rm \ at \ fixed\/}\ f}
E(|\Phi\rangle\langle\Phi|)=
\frac{1}{2}\left({1-\sqrt{1-f^2}}\right).  
\end{equation}
Thus, as a function of $f$, 
$\epsilon(f)$ is given by 
\begin{equation}
\epsilon(f)=
\begin{cases}
\frac{1}{2}\left({1-\sqrt{1-f^2}}\right)&{\rm for} \ f\le 0,\cr
0&{\rm for} \ f\ge 0, 
\end{cases}
\end{equation}
which, being convex for $f\in[-1,1]$, gives 
the entanglement function~(\ref{eqn:Werner}) for Werner states. 

\subsection{Isotropic states}
These are states invariant under 
\begin{equation}
{\rm \bf P}_{2}:\rho\rightarrow 
\int dU\,(U\otimes U^*)\rho\,
(U^\dagger\otimes {U^*}^\dagger),
\end{equation} and can be expressed as 
\begin{eqnarray}
\rho_{\rm iso}(F)
\equiv
\frac{1-F}{d^2-1}
\left(\hat{\openone}-|\Phi^+\rangle\langle\Phi^+|\right)+
F|\Phi^+\rangle\langle\Phi^+|,
\end{eqnarray}
where $|\Phi^+\rangle\equiv\frac{1}{\sqrt{d}}\sum_{i=1}^{d}|ii\rangle$ and
$F\in[0,1]$. For $F\in[0,1/d]$ this state is known to be separable~\cite{Horodecki299}.
By following 
arguments similar to those applied by 
Terhal and Vollbrecht~\cite{TerhalVollbrecht00}
for $E_{\rm F}(\rho_{\rm iso})$ one arrives 
at 
\begin{eqnarray}
\label{eqn:Eiso}
 \!\!\!\!\!\! E_{\sin^2}\left(\rho_{\rm iso}(F)\right)=
1-\frac{1}{d}\big(\sqrt{{F}}
+\sqrt{(\!1-\!F)(d\!-\!1)}\,\big)^2,
\end{eqnarray}
for $F\ge 1/d$.
The essential point of the derivation is the following Lemma (cf.~Ref.~\cite{TerhalVollbrecht00}): \\
{\it Lemma 1\/}. The entanglement $E_{\sin^2}$ for isotropic states in $C^d\otimes
C^d$ for $F\in[1/d,1]$ is given by
\begin{equation}
E_{\sin^2}(\rho_{\rm iso}(F))={\rm co\/}(R(F)),
\end{equation}
where ${\rm co}(R(F))$ is the convex hull of the function $R$ and
\begin{equation}
R(F)=1-\max_{\{\mu_i\}}\big\{\mu_i\,|F=\Big(\sum_{i=1}^d\sqrt{\mu_i}\Big)^2/d
;\ \sum_{i=1}^d \mu_i=1\big\}.
\end{equation}
Straightforward extremization shows that 
\begin{equation}
R(F)=1-\left(\sqrt{\frac{F}{d}}+\sqrt{\frac{F+d-1}{d}-F}\right)^2,
\end{equation}
which is convex, and hence ${\rm co}(R(F))=R(F)$. Thus we arrive at the entanglement result for isotropic states given in Eq.~(\ref{eqn:Eiso}).

\begin{figure}[t]
\centerline{\psfig{figure=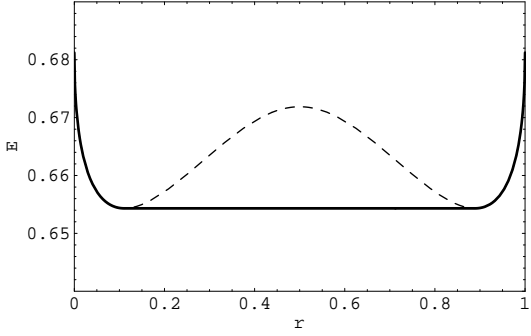,width=7cm,height=4.5cm,angle=0}}
%\vspace{- 0.4cm}
\caption{Entanglement curve for the mixed state
$\rho_{7;2,5}(r)$ (full line) 
constructed as the convex hull of the curve for the pure state 
$\ket{{\rm S}{\rm S}_{7;2,5}(r,\phi)}$ 
(dashed in the middle; full at the edges).}
\label{fig:SS72}
%\vspace{-0.5cm}
\end{figure}

\subsection{Mixtures of multipartite symmetric states}
\label{sec:MultiMixed}
Before exploring more general mixed states, it 
is useful to first examine states with high symmetry.
With this in mind, we consider states formed by mixing two
distinct symmetric states (i.e., $k_1\ne k_2$):
\begin{eqnarray}
&&\rho_{n;k_1\!k_2}\!(r)\equiv r\, 
\ket{{\rm S}(n,k_1)}\bra{{\rm S}(n,k_1)}\nonumber \\
&&\qquad\qquad+(1-r)\ket{{\rm S}(n,k_2)}\bra{{\rm S}(n,k_2)}.
\end{eqnarray}
From the independence of 
$E_{\sin^2}\left(\ket{{\rm S}{\rm S}_{n;k_1k_2}(r,\phi)}\right)$ 
on $\phi$ and the fact that the mixed state $\rho_{n;k_1\!k_2}\!(r)$
is invariant under the projection 
\begin{equation}
{\rm P_3}:\rho\rightarrow \int\frac{d\phi}{2\pi}U^{\otimes n}\rho\, U^{\dagger\otimes n}
\end{equation}
with $U:\big\{\ket{0},\ket{1}\big\}\to
\big\{\ket{0},{\rm e}^{-i\phi}\ket{1}\big\}$, we have that 
$E_{\sin^2}\left(\rho_{n;k_1k_2}(r)\right)$ vs.~$r$ can be 
constructed from the convex hull of the entanglement function of 
$\ket{{\rm S}{\rm S}_{n;k_1k_2}(r,0)}$ vs.~$r$.  An example, $(n,k_1,k_2)=(7,2,5)$, is 
shown in Fig.~\ref{fig:SS72}. 
If the dependence of $E_{\sin^2}$ on $r$ is already convex for the 
pure state, its mixed-state counterpart has precisely the same dependence.  
Figure~\ref{fig:WW}, for which $(n,k_1,k_2)=(3,1,2)$, 
exemplifies such behavior. More generally, one can consider mixed states
of the form 
\begin{equation}
\label{eqn:mixSnk}
\rho(\{p\})=\sum_k p_k\ketbra{S(n,k)}.
\end{equation}
The entanglement $E_{\rm mixed}(\{p\})$ can then be obtained as a function of the mixture $\{p\}$ from the convex hull of the entanglement function $E_{\rm pure}(\{q\})$ for the {\it pure\/} state $\sum_k\sqrt{q_k}\ket{S(n,k)}$. 
That is, $E_{\rm mix}(\{p\})={\rm co}\,E_{\rm pure}(\{q\}=\{p\})$.
Therefore, the entanglement for a mixture of symmetric states $\ket{S(n,k)}$
is known, up to some convexification.

\section{Application to arbitrary mixture of GHZ, W and inverted-W states}
\label{sec:GWW}
Having warmed up in Sec.~\ref{sec:MultiMixed} by analyzing mixtures of
multipartite symmetric states, we now turn our attention to mixtures of
three-qubit GHZ, W and inverted-W states.
\subsection{Symmetry and entanglement preliminaries}
These states are 
important, in the sense that all pure states can,
under stochastic LOCC, be transformed either to GHZ or W (equivalently inverted-W) states. It is thus interesting to determine the entanglement
content (using any measure of entanglement) for mixed states of the form:
\begin{equation}
\label{eqn:GWW}
\rho(x,y)\equiv x\ketbra{\rm GHZ}+y\ketbra{\rm W}+(1-x-y)\ketbra{\widetilde{\rm W}},
\end{equation}
where $x,y\ge 0$ and $x+y\le 1$.  This family of mixed states
is not contained in the family~(\ref{eqn:mixSnk}), 
as $\ket{\rm GHZ}=\big(\ket{S(3,0)}+\ket{S(3,3)})/\sqrt{2}$.
The property of $\rho(x,y)$ that facilitates 
the computation of its entanglement is a certain invariance, which 
we now describe.  Consider the local unitary transformation on a single 
qubit: 
\begin{subequations}
\begin{eqnarray}
\ket{0}&\rightarrow& \ket{0}, \\
\ket{1}&\rightarrow& g^k\ket{1}, 
\end{eqnarray}
\end{subequations}
where $g=\exp{(2\pi i/3)}$, i.e., a relative phase shift.  
This transformation, when applied simultaneously to all 
three qubits, is denoted by $U_k$.  It is straightforward 
to see that $\rho(x,y)$ is invariant under the mapping 
\begin{equation}
{\rm\bf P}_4:\rho\rightarrow 
\frac{1}{3}\sum_{k=1}^3
U_k\,\rho\,U_k^\dagger\,.
\end{equation}
Thus, we can apply Vollbrecht-Werner 
technique~\cite{VollbrechtWerner01} 
to the compution of the entanglement of $\rho(x,y)$.

\begin{figure}[t]
\centerline{\psfig{figure=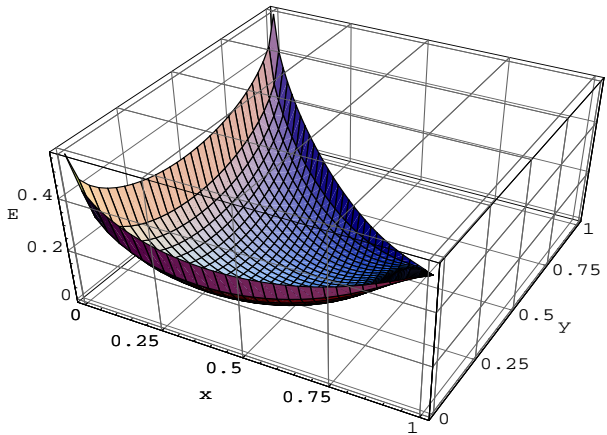,width=8cm,height=6cm,angle=0}}
%\vspace{- 0.4cm}
\caption{Entanglement vs.~the composition of the pure state 
$\ket{\psi(x,y)}$. This entanglement surface is not convex near $(x,y)=(1,0)$,
although not obvious from the plot.}
\label{fig:gGWW0}
%\vspace{-0.2cm}
\end{figure} 
Now, the Vollbrecht-Werner procedure requires one to characterize
the set $S_{\rm inv}$ of all pure states that are invariant under the
projection ${\rm\bf P}_4$. Then, the convex hull of $E_{\sin^2}(\rho)$ need only be
taken over $S_{\rm inv}$, instead of the set of {\it all\/} pure states.
However, as the state $\rho(x,y)$ is a mixture of three orthogonal
pure states ($\ket{\rm GHZ}$, $\ket{\rm W}$ and $\ket{\widetilde{\rm W}}$) that
are themselves invariant under ${\rm\bf P}_4$, the
pure states that can enter any possible decomposition of
$\rho$ must be of the restricted form:
\begin{equation}
\alpha \ket{\rm GHZ}+\beta\ket{\rm W}+\gamma\ket{\widetilde{\rm W}},
\end{equation} 
with $|\alpha|^2+|\beta|^2+|\gamma|^2=1$. Thus, there is no
need to characterize $S_{\rm inv}$, but rather to characterize
the pure states that, under ${\rm {\bf P}_4}$, are projected to $\rho(x,y)$.  
These states are readily seen to be of the form:
\begin{equation}
\sqrt{x}\,e^{i\phi_1}\ket{\rm GHZ}+
\sqrt{y}\,e^{i\phi_2}\ket{\rm W}+
\sqrt{1-x-y}\,e^{i\phi_3}\ket{\widetilde{\rm W}}.
\end{equation}
Of these, the least entangled state, for $(x,y)$ given, has all 
coefficients non-negative (up to a global phase), i.e., 
\begin{equation}
\ket{\psi(x,y)}\equiv
\sqrt{x}\ket{\rm GHZ}+
\sqrt{y}\ket{\rm W}+
\sqrt{1-x-y}\ket{\widetilde{\rm W}}.
\end{equation}
The entanglement eigenvalue of $\ket{\psi(x,y)}$ can then be readily 
calculated, and one obtains
\begin{equation}
\Lambda(x,y)\!=\!\frac{1}{(1 \!+\! t^2)^{\frac{3}{2}}} \left\{\sqrt{\frac{x}{2}}(1\! +\! t^3) + \sqrt{3y}\,t + 
      \sqrt{3(1\!-\!x\!-\!y)}\,t^2\right\}, 
\end{equation}
where $t$ is the (unique) non-negative real root of the following 
third-order polynomial equation:
\begin{eqnarray}
&&3\sqrt{\frac{x}{2}}(-t + t^2) + \sqrt{3y}(-2t^2 + 1)
\nonumber\\ 
&&\qquad\qquad\quad
+\sqrt{3(1 - x-y)}(-t^3 + 2t) = 0.
\end{eqnarray}
Hence, the entanglement function for $\ket{\psi(x,y)}$, i.e., 
$E_\psi(x,y)\equiv 1-\Lambda(x,y)^2$, 
is determined (up to the straightforward task of root-finding).
\begin{figure}%[h]
\centerline{\psfig{figure=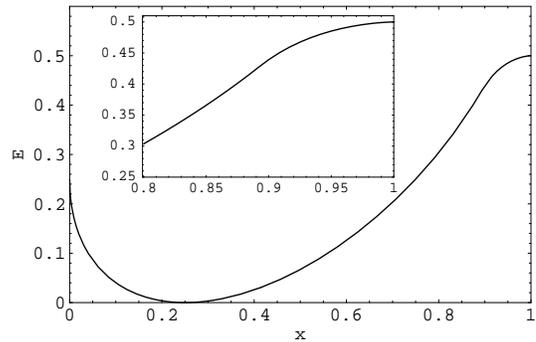,width=7cm,height=4.5cm,angle=0}}
%\vspace{- 0.4cm}
\caption{Entanglement of the pure state 
$\ket{\psi\big(x,y=(1-x)/2\big)}=\sqrt{x}\,\ket{\rm GHZ}+\sqrt{(1-x)/2}\,\ket{W}+\sqrt{(1-x)/2}\,\ket{\widetilde{W}}$ 
vs.~$x$.
This shows the entanglement along the diagonal boundary $x+2y=1$.  
Note the absence of convexity near $x=1$; this region is 
repeated in the inset.}  
\label{fig:gGW}
%\vspace{-0.5cm}
\end{figure}

\subsection{Finding the convex hull}
Recall that our aim is to determine the entanglement of the mixed 
state $\rho(x,y)$.  As we already know the entanglement of the 
corresponding pure state $\ket{\psi(x,y)}$, we may accomplish our aim 
by the Vollbrecht-Werner technique~\cite{VollbrechtWerner01}, 
which gives the entanglement of $\rho(x,y)$ in terms of that 
of $\ket{\psi(x,y)}$ via the convex hull construction: 
$E_\rho(x,y)=({\rm co}\,E_\psi)(x,y)$.  Said in words, the entanglement surface 
$z=E_{\rho}(x,y)$ is the convex surface constructed from the surface 
$z=E_{\psi}(x,y)$.  

The idea underlying the use of the convex hull is this.  Due to its 
linearity in $x$ and $y$, the state $\rho(x,y)$~(\ref{eqn:GWW}) 
can [except when $(x,y)$ lies on the boundary] be decomposed into 
two parts: 
\begin{equation}
\rho(x,y)=p \,\rho(x_1,y_1) +(1-p)\rho(x_2,y_2),
\end{equation} 
with the weight $p$ and end-points $(x_1,y_1)$ and $(x_2,y_2)$ 
related by 
\begin{subequations}
\begin{eqnarray}
p\, x_1+(1-p)x_2=x \\ 
p\, y_1+(1-p)y_2=y.
\end{eqnarray}
\end{subequations}
Now, if it should happen that 
\begin{equation}
p   E_{\psi}(x_1,y_1)+
(1-p)E_{\psi}(x_2,y_2)< 
     E_{\psi}(x,y)
\end{equation}
then the entanglement, averaged over the end-points, would give a value 
lower than that at the interior point $(x,y)$; this conforms with 
the convex-hull construction.

It should be pointed out that the convex hull should be taken 
with respect to parameters on which the density matrix depends 
{\it linearly\/}, such as $x$ and $y$ in the example of $\rho(x,y)$.  
Furthermore, in order to obtain the convex hull of a function, 
one needs to know the {\it global\/} structure of the function---in the present case, $E_{\psi}(x,y)$.  We note that numerical 
algorithms have been developed for constructing convex 
hulls~\cite{QHull}. 
\begin{figure}[t]
\centerline{\psfig{figure=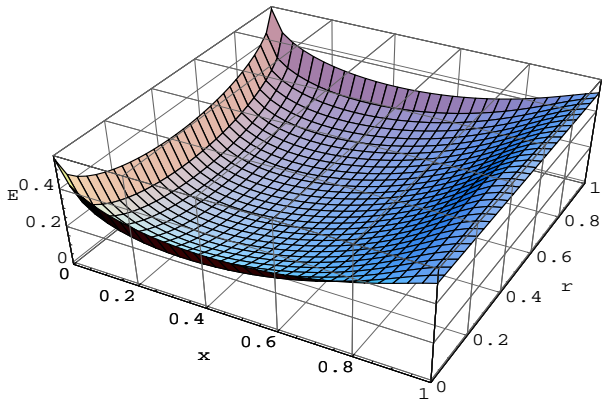,width=8cm,height=5.5cm,angle=0}}
%\vspace{- 0.4cm}
\caption{Entanglement of the pure state 
$\ket{\psi\big(x,(1-x)r\big)}=\sqrt{x}\,
\ket{\rm GHZ}+
\sqrt{(1-x)r}\,\ket{W}+
\sqrt{(1-x)(1-r)}\ket{\widetilde{W}}$ vs.~$x$ and $r$.  
Note the symmetry of the surface with respect with $r=1/2$.}  
\label{fig:gGWWxr0}
%\vspace{-0.2cm}
\end{figure}
\begin{figure}[t]
\centerline{\psfig{figure=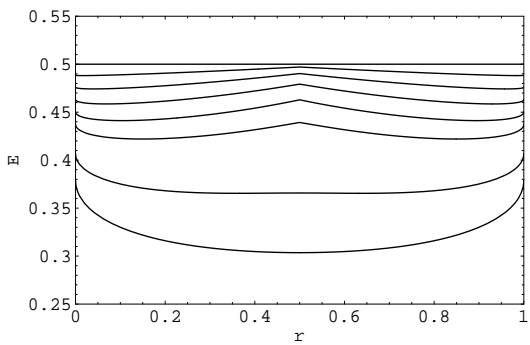,width=7cm,height=4.5cm,angle=0}}
%\vspace{- 0.4cm}
\caption{Entanglement of the pure states
$\ket{\psi\big(x,(1-x)r\big)}=
\sqrt{x}\,
\ket{\rm GHZ}+
\sqrt{(1-x)r}\,\ket{W}+
\sqrt{(1-x)(1-r)}\ket{\widetilde{W}}$
vs.~$r$ for various values of $x$ 
(from the bottom: 0.8, 0.85, 0.9, 0.92, 0.94, 0.96, 0.98, 1). 
This reveals the nonconvexity in $r$ for intermediate values of $x$.}  
\label{fig:gGWWxr1}
%\vspace{-0.2cm}
\end{figure}

As we have discussed, our route to establishing the entanglement of 
$\rho(x,y)$ involves the analysis of the entanglement of $\ket{\psi(x,y)}$, 
which we show in Fig.~\ref{fig:gGWW0}.  Although it is not obvious, 
the corresponding surface fails to be convex near the point $(x,y)=(1,0)$, 
and; therefore in this region we must suitably convexify in order to 
obtain  the entanglement of $\rho(x,y)$.  
To illustrate the properties 
of the entanglement of $\ket{\psi(x,y)}$ we show, in Fig.~\ref{fig:WW}, 
the entanglement of $\ket{\psi(x,y)}$ along the line $(x,y)=(0,s)$; 
evidently this is convex.  By contrast, along the line $x+2y=1$ 
there is a region in which the entanglement is not convex, as 
Fig.~\ref{fig:gGW} shows.  The nonconvexity of the entanglement of 
$\ket{\psi(x,y)}$ complicates the calculation of the entanglement of 
$\rho(x,y)$, as it necessitates a procedure for constructing the 
convex hull in the (as it happens, small) nonconvex region.
Elsewhere in the $xy$ plane the entanglement of $\rho(x,y)$ is given 
directly by the entanglement of $\ket{\psi(x,y)}$.  

At worst, convexification would have to be undertaken numerically.  
However, in the present setting it turns out that one can determine the 
convex surface essentially analytically, by performing the necessary 
surgery on surface $z=E_{\psi}(x,y)$.  To do this, we make use of the 
fact that if we parametrize $y$ via $(1-x)r$, i.e., we consider 
\begin{eqnarray}
{\rho\big(x,(1-x)r\big)}&=&x\,
\ketbra{\rm GHZ}+(1-x)r\,\ketbra{W}\nonumber \\
&&+(1-x)(1-r)\ketbra{\widetilde{W}}, 
\label{eqn:Rhoxr}
\end{eqnarray}
where $0\le r\le 1$ [and similarly for $\ket{\psi(x,y)}$] then, 
as a function of $(x,r)$, the entanglement will be symmetric with 
respect to $r=1/2$, as Fig.~\ref{fig:gGWWxr0} makes evident. 
With this parametrization, the nonconvex region of the entanglement 
of $\ket{\psi}$ can more clearly be identified.  

\begin{figure}[t]
\centerline{\psfig{figure=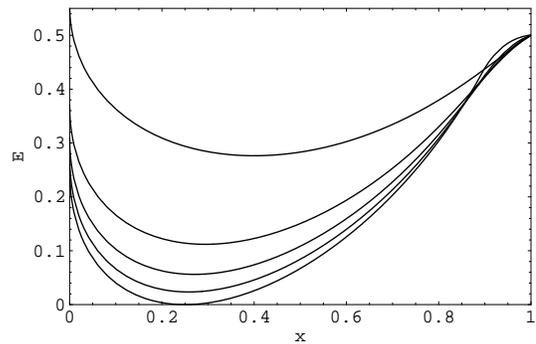,width=7cm,height=4.5cm,angle=0}}
%\vspace{- 0.2cm}
\caption{Entanglement of the pure states 
$\ket{\psi\big(x,(1-x)r\big)}=
\sqrt{x}\,\ket{\rm GHZ}+
\sqrt{(1-x)r}\,\ket{W}+
\sqrt{(1-x)(1-r)}\ket{\widetilde{W}}$ vs.~$x$ 
for various values of $r$ 
(from the top: 0, 0.1, 0.2, 0.3, 0.5). 
This reveals the nonconvexity in $x$ in the (approximate) 
interval $[0.85,1]$.}  
\label{fig:gGWWxr2}
%\vspace{-0.4cm}
\end{figure}
To convexify this surface we adopt the following convenient strategy. 
First, we reparametrize the coordinates, exchanging $y$ by $(1-x)r$.  
Now, owing to the linearity, in $r$ at fixed $x$ and vice versa, of the 
coefficients $x$, $(1-x)r$ and $(1-x)(1-r)$ in Eq.~(\ref{eqn:Rhoxr}), 
it is certainly necessary for the entanglement of $\rho$ to be a 
convex function of $r$ at fixed $x$ and vice versa.  Convexity is, 
however, not necessary in other directions in the $(x,r)$ plane, owing to 
the nonlinearity of the the coefficients under simultaneous variations 
of $x$ and $r$.  Put more simply: convexity is not necessary throughout 
the $(x,r)$ plane because straight lines in the $(x,r)$ plane 
do not correspond to straight lines in the $(x,y)$ plane
(except along lines parallel either to the $r$ or the $x$ axis).
Thus, our strategy will be to convexify in a restricted sense: 
first along lines parallel to the $r$ axis and then along lines 
parallel to the $x$ axis.  Having done this, we shall check to see 
that no further convexification is necessary.

For each $x$, we convexify the curve $z=E_{\psi}\big(x,(1-x)r\big)$ as 
a function of $r$, and then generate a new surface by allowing $x$ to 
vary.  More specifically, the nonconvexity in this direction has the 
form of a symmetric pair of minima located on either side of a cusp, 
as shown in Fig.~\ref{fig:gGWWxr1}.  Thus, to correct for it, we 
simply locate the minima and connect them by a straight line. 

What remains is to consider the issue of convexity along the $x$ 
(i.e., at fixed $r$) direction for the surface just constructed.  
In this direction, nonconvexity occurs when $x$ is, roughly speaking, 
greater than $0.8$, as Fig.~\ref{fig:gGWWxr2} suggests.  In contrast 
with the case of nonconvexity at fixed $r$, this nonconvexity is due 
to an inflection point at which the second derivative vanishes.
To correct for it, we locate the point $x=x_0$ such that the tangent 
at $x=x_0$ is equal to that of the line between the point on the curve 
at $x_0$ and the end-point at $x=1$, and connect them with a straight 
line.  This furnishes us with a surface convexified with respect to 
$x$ (at fixed $r$) and vice versa. 

\begin{figure}[t]
\centerline{\psfig{figure=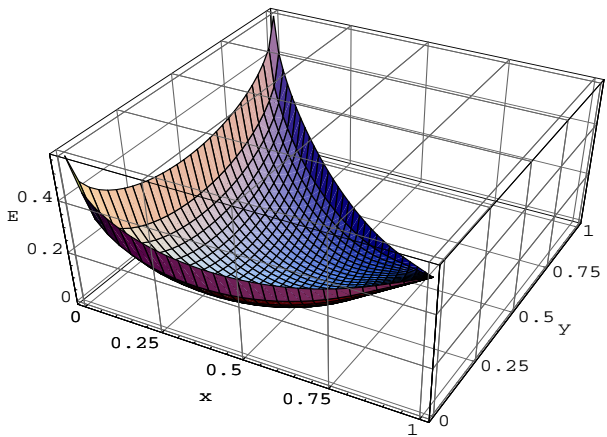,width=8cm,height=6cm,angle=0}}
%\vspace{- 0.2cm}
\caption{Entanglement of the mixed state ${\rho(x,y)}$.}
\label{fig:gGWW}
%\vspace{-0.5cm}
\end{figure} 
Armed with this surface, we return to the $(x,y)$ parametrization, 
and ask whether or not it is fully convex (i.e., convex along straight 
lines connecting {\it any\/} pair of points).  Said equivalently, we 
ask whether or not any further convexification is required.  Although 
we have not proven it, on the basis of extensive numerical exploration 
we are confident that the resulting surface is, indeed, convex.  The 
resulting convex entanglement surface for $\rho(x,y)$ is shown in 
Fig.~\ref{fig:gGWW}.  Figure~\ref{fig:gGWW05} exemplifies this 
convexity along the line $2y+x=1$.  We have observed that for the 
case at hand it is adequate to correct for nonconvexity only in the 
$x$ direction at fixed $r$.
\begin{figure}[t]
\centerline{\psfig{figure=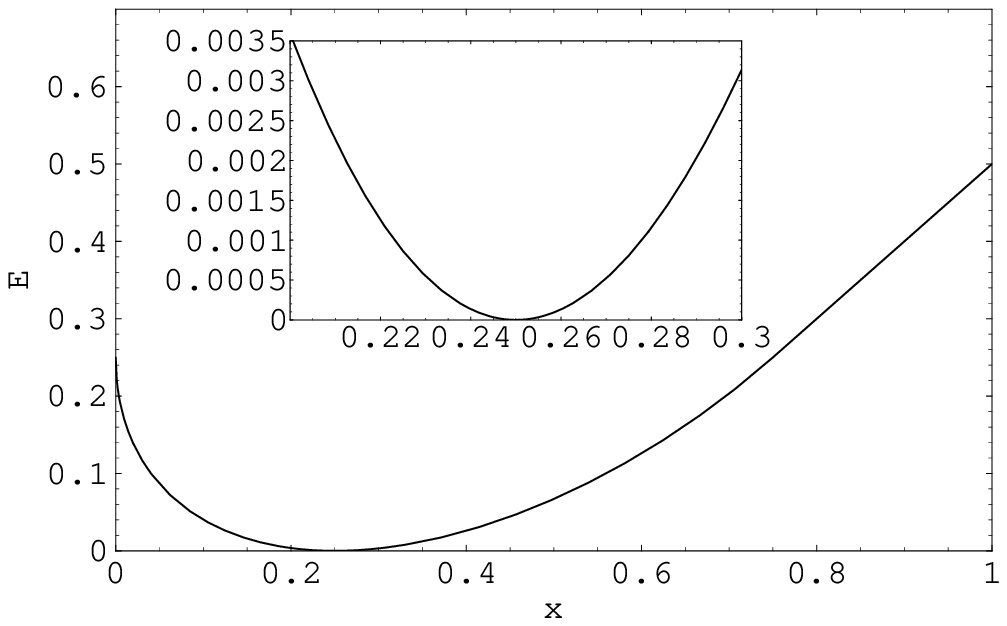,width=7cm,height=4.5cm,angle=0}}
%\vspace{- 0.4cm}
\caption{Entanglement of the mixed state 
${\rho\big(x,y=(1-x)/2\big)}=x\,
\ketbra{\rm GHZ}+
\frac{1-x}{2}\big(\ketbra{W}+
\ketbra{\widetilde{W}}\big)$ vs.~$x$.  
Inset: enlargement of the region $x\in[0.2,0.3]$. 
This contains the only point, $(x,y)=(1/4,3/8)$, 
at which $E_{\rho}(x,y)$ vanishes.}  
\label{fig:gGWW05}
%\vspace{-0.4cm}
\end{figure}

\subsection{Comparison with the negativity}
This measure of entanglement is defined to be twice the absolute value 
of the sum of the negative eigenvalues of the partial transpose of 
the density
matrix~\cite{ZyczkowskiWerner,WeiNemotoGoldbartKwiatMunroVerstraete03}.
In the present setting, 
viz., the family $\rho(x,y)$ of three-qubit states, the partial 
transpose may equivalently be taken with respect to any one of the 
three parties, owing to the invariance of $\rho(x,y)$ under all 
permutations of the parties.  Transposing with respect to the third 
party, one has 
\begin{equation}
N(\rho)\equiv-2\sum_{\lambda_i<0} \lambda_i, 
\end{equation}
where the $\lambda$'s are the eigenvalues of the matrix $\rho^{T_3}$, 

It is straightforward to calculate the negativity of $\rho(x,y)$; 
the results are shown in Fig.~\ref{fig:Nxy}.  
Interestingly, for all allowed
values of $(x,y)$, the state $\rho(x,y)$ has nonzero negativity, except
at $(x,y)=(1/4,3/8)$, at which the calculation of the GME shows that 
the density matrix is indeed separable. One also sees that $\rho(1/4,3/8)$
is a separable state from that fact that it can be obtained by applying
the projection ${\rm \bf P}_4$ to the (un-normalized) separable pure state 
$\big(\ket{0}+\ket{1}\big)^{\otimes 3}$. 
The fact that the only 
positve-partial-transpose (PPT)
state is separable is the statement that there are no entangled PPT states 
(i.e., no PPT bound entangled states) within this family of three-qubit 
mixed states.  The negativity surface, Fig.~\ref{fig:Nxy}, is 
qualitatively---but not quantitatively---the same as that of GME. 
By inspecting the negativity and GME surfaces one can see that they 
present ordering difficulties.  This means that one can find 
pairs of states $\rho(x_1,y_1)$ and $\rho(x_2,y_2)$ that have respective 
negativities $N_1$ and $N_2$ and GMEs $E_1$ and $E_2$ such 
that, say, $N_1< N_2$ but $E_1>E_2$.  Said equivalently, the negativity 
and the GME do not necessarily agree on which of a pair of states is 
the more entangled.  For two qubit settings, such ordering difficulties 
do not show up for pure states but can for mixed states~\cite{Ordering,WeiNemotoGoldbartKwiatMunroVerstraete03}.  
On the other hand, for three qubits, such ordering difficulties already show up for pure states, 
as the following example shows: $N({\rm GHZ})=1>N({\rm W})=2\sqrt{2}/3$ 
whereas for the GME the order is reversed.  We note, however, that for the 
relative entropy of entanglement $E_R$, one has $E_R({\rm GHZ})=\log 2 < E_R({\rm W})=\log(9/4)$~\cite{PlenioVedral01}, which for this
particular case is in accord with the GME.

\begin{figure}[t]
%\vspace{0.2cm}
\centerline{\psfig{figure=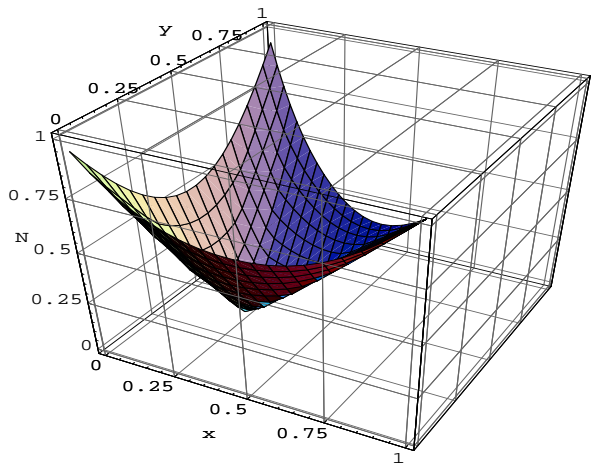,width=7.8cm, height=6.2cm,angle=0}}
%\vspace{- 0.3cm}
\caption{Negativity of the mixed state ${\rho(x,y)}$.}
\label{fig:Nxy}
%\vspace{-0.5cm}
\end{figure}

\section{Concluding remarks}
\label{sec:Conclude}
We have considered a rather general, geometrically motivated, measure 
of entanglement, applicable to pure and mixed quantum states involving 
arbitrary numbers and structures of parties. In bipartite settings, this 
approach provides an alternative---and generally inequivalent---measure to the entanglement 
of formation. For multipartite settings, there is, to date, no explicit generalization of entanglement of formation~\cite{PlenioVedral01}. However, if such a generalization 
should emerge, and if it should be based on the convex hull construction 
(as it is in the bipartite case), then one may be able to calculate the 
entanglement of formation for the families of multipartite mixed states 
considered in the present Paper.  

As for explicit implementations, the geometric measure of entanglement yields analytic results in several bipartite cases for which the entanglement of formation is already known. These cases include: (i)~arbitrary two-qubit mixed, 
(ii)~generalized Werner, and (iii)~isotropic states. 
Furthermore, we have obtained
the geometric measure of entanglement for certain multipartite mixed states, such as mixtures of symmetric states.
In addition, by making use of the geometric measure, we have addressed 
the entanglement of a rather general family of three-qubit mixed 
states analytically (up to root-finding).  This family consists 
of arbitrary mixtures of GHZ, W, and inverted-W states.  To the 
best of our knowledge, corresponding results have not, to date, 
been obtained for other measures of entanglement, such as 
entanglement of formation and relative entropy of entanglement.  
We have also obtained  corresponding results for the 
negativity measure of entanglement. Among other 
things, we have found that there are no PPT bound entangled states 
within this general family.

A significant issue that we have not discussed is how to use the geometric measure to provide a classification of entanglement of various 
multipartite entangled states, even in the pure-state setting. 
For example, given a tripartite state, is all the entanglement associated
with pairs of parts, or is some attributable only to the system as a whole? 
More generally, one can envisage all possible partitionings of the parties, 
and for each, compute the geometric measure of entanglement.
This would provide a hierarchical characterization of the entanglement
of states, more refined than
the global characterization discussed here. Another extension would involve augmenting the set of separable
pure states with certain classes of entangled 
pure states, such as biseparable entangled, W-type and GHZ-type states~\cite{AcinBrussLewensteinSanpera01}. 

Although there is no generally valid analytic procedure for
computing the entanglement eigenvalue $\Lambda_{\max}$, one can give---and indeed we have given---analytical results for several elementary cases.  Harder
examples require computation, but
often this is (by today's computational standards) trivial.
We note that in order to find $\Lambda_{\max}$ for the state
$\ket{\psi}$ it is not necessary
to solve the nonlinear eigenproblem~(\ref{eqn:Eigen}); one can instead
appropriately parameterize the family of separable states $\ket{\phi}$ and then
directly maximize
their overlap with the entangled state $\ket{\psi}$, i.e., $\Lambda_{\max}=\max_\phi||\ipr{\phi}{\psi}||$. 
As an aside, we mention there exist numerical techniques for determing $E_{\rm F}$
(see, e.g., Ref.~\cite{Zyczkowski99}). We believe that numerical techniques
for solving the geometric measure of entanglement for general multipartite
mixed states can readily be developed. 

The motivation for constructing the measure discussed
in the present Paper is that we wish to address the degree of entanglement from
a geometric viewpoint, regardless of the number of parties.
Although the construction is purely geometric, we have related this measure
to entanglement witnesses, which can in principle be measured 
locally~\cite{Guhne}. Moreover, the geometric measure of entanglement
is related to the probability of preparing a single copy of a two-qubit mixed state from a certain pure state~\cite{Vidalb00}. 
Yet it is still desirable to see whether, in general, this measure can be
associated with 
any physical process in quantum information, as are the entanglement
of formation and distillation.

There are further issues that remain to be explored, such as additivity and ordering.
The present form of entanglement for pure states, $E_{\sin^2}\equiv 1-\Lambda^2$, is not additive.
However, one can consider a related form, $E_{\log}\equiv -\log\Lambda^2$,
which, e.g., is additive for $\ket{\psi}_{AB}\otimes\ket{\psi}_{CD}$, i.e.,
\begin{equation}
E_{\log}\big( \ket{\psi_1}_{AB}\otimes\ket{\psi_2}_{CD}\big)=
E_{\log}\big(\ket{\psi_1}_{AB}\big)+E_{\log}\big(\ket{\psi_2}_{CD}\big).
\end{equation}
This suggests that it is more appropriate to use this logarithmic form of entanglement 
to discuss additivity issues. However, it remains to check whether it
is an entanglement monotone when extended to mixed states by convex hull.

As regards the ordering issue, we first mention a result of bipartite entanglement measures, due to Virmani 
and Plenio~\cite{Ordering}, which states that any two measures with continuity 
that give the same value as the entanglement of formation for {\it pure\/} states are either \lq\lq identical or induce different orderings in general\rlap.\rq\rq\
This result points out that
different entanglement measures will inevitably induce different
orderings if they are inequivalent.
This result might still hold for multipartite settings, despite 
their discussion 
being based on the existence of entanglement of formation and 
distillation, which have not been generalized to multipartite settings. 
Although the geometric measure gives the same ordering as 
the entanglement of formation for two-qubit mixed states [see 
Eq.~(\ref{eqn:EC})], we belive that the geometric measure 
will, in general, give a different ordering.  However, it is not our intention to discuss the ordering difficulty in the present Paper.  Nevertheless, 
it is 
interesting to point out that for bipartite systems, even though the 
relative entropy of entanglement coincides with entanglement of 
formation for pure states, they can give different orderings for mixed
states, as pointed out by Verstraete 
et al.~\cite{Ordering}. 

We conclude by remarking that the measure discussed in the present Paper is not 
included among the infinitely many different measures proposed by Vedral 
et al.~\cite{VedralPlenioRippinKnight97}.  Those measures are based on the minimal distance between the entangled mixed state and the set of separable {\it mixed\/} states. 
By contrast, the measure considered here is based upon the minimal 
distance between the entangled pure state and the set of separable pure states,
and it is extended to mixed states
by a convex hull construction. 

\section*{Acknowledgements}
We thank J.~Altepeter, H.~Barnum, D.~Bru\ss, W.~D\"ur, H.~Edelsbrunner, J.~Eisert, A.~Ekert, M.~Ericsson, O.~G\"uhne, L.-C.~Kwek, P.~Kwiat, D.~Leung, C.~Macchiavello, S.~Mukhopadhyay, Y.~Omar,
M.~Randeria, F.~Verstraete, G.~Vidal and especially W.~J.~Munro for discussions.
PMG acknowledges the hospitality of the 
Unversity of Colorado--Boulder and the Aspen Center for Physics. 
This work was supported by 
NSF EIA01-21568 and 
DOE DEFG02-91ER45439.
TCW acknowledges a Mavis Memorial Fund Scholarship and a Harry G. Drickamer
Graduate Award. TCW also acknowledges the hospitality of the Benasque Center 
for Science, where part of the revision was done.

\bigskip
\appendix
\section{The Vollbrecht-Werner technique}
\label{app:VW}
In this Appendix, we now briefly review a technique 
developed by Vollbrecht and Werner~\cite{VollbrechtWerner01} for 
computing the entanglement of formation for the generalized Werner
states; this 
turns out to be applicable to the computation of the sought 
quantity $E_{\sin^2}$.  We start by fixing some notation.  
Let 
\begin{itemize}
\item[(a)] $K$ be a compact convex set 
(e.g., a set of states that includes both pure and mixed ones); 
\item[(b)] $M$ be a convex subset of $K$ (e.g., set of pure states); 
\item[(c)] $E:M\rightarrow R\cup\{+\infty\}$ be a function that maps 
elements of $M$ to the real numbers (e.g., $E=E_{\sin^2}$); and 
\item[(d)] $G$ be a compact group of symmetries, acting on $K$
(e.g., the group $U\otimes U^\dagger$) as 
$\alpha_g:K\rightarrow K$ 
(where $\alpha_g$ is the representation of the element $g\in G$)
that preserve convex combinations.
\end{itemize}

We assume that 
$\alpha_g M\subset M$ 
(e.g., pures states are mapped into pure states), 
and that 
$E(\alpha_g m)=E(m)$ for all $m\in M$ and $g\in G$ 
(e.g., that the entanglement of a pure state is preserved under 
$\alpha_g$).  
We denote by ${\rm\bf P}$ the invariant projection operator 
defined via 
\begin{equation}
{\rm\bf P}k=
\int dg\,\alpha_g(k),
\end{equation}
where $k\in K$.  
Examples of ${\rm\bf P}$ are the operations ${\rm\bf P}_1$ and ${\rm\bf P}_2$ in the main text.
Vollbrecht and Werner also defined the following real-valued function 
$\epsilon$ on the invariant subset ${\rm\bf P}K$: 
\begin{equation}
\epsilon(x)=
{\rm inf}
\left\{E(m)\vert m\in M, {\rm \bf P}m=x\right\}.
\end{equation}

They then showed that, for $x\in{\rm\bf P}K$,
\begin{equation}
{\rm co}\,E(x)=
{\rm co}\,\epsilon(x),
\end{equation}
and provided the following recipe for computing the function 
${\rm co}\,E$ for
$G$-invariant states:
\begin{itemize}
\item[1.] For every invariant state $\rho$ 
(i.e., obeying $\rho={\rm\bf P}\rho$),  
find the set $M_\rho$ of pure states
$\sigma$ such that ${\rm \bf P}\sigma=\rho$.
\item[2.] Compute 
$\epsilon(\rho)\equiv{\rm inf}
\left\{E(\sigma)\vert \sigma\in M_\rho\right\}$.
\item[3.] Then ${\rm co}\,E$ is the convex hull of this function $\epsilon$. 
\end{itemize}

\end{document}